\documentclass[10pt]{iopart}

\usepackage{graphicx}
\begin{document}

\title[Universal Crossover in Improved Perturbation Theory]{Asymptotically Universal Crossover in Perturbation Theory with a Field Cutoff}

\author{L. Li\dag\ and Y. Meurice\dag\ \S
\footnote[3]{To
whom correspondence should be addressed (yannick-meurice@uiowa.edu)}
}
\address{\dag\ Department of Physics and Astronomy\\ The University of Iowa\\
Iowa City, IA 52242 USA }
\address{\S\ Also at the Obermann Center for Advanced Study, University of Iowa}
\date{\today}

\begin{abstract}
We discuss the crossover between the small and large field cutoff (denoted $x_{max}$) limits of the perturbative coefficients for a simple integral and the anharmonic oscillator. 
We show that in the limit where the order $k$ of the perturbative coefficient $a_k(x_{max})$ becomes large and for $x_{max}$ in the crossover region, $a_k(x_{max})\propto \int_{-\infty}^{x_{max}}{\rm e}^{-A(x-x_0(k))^2}dx$. The constant $A$ and the function $x_0(k)$ are determined 
empirically and compared with exact (for the integral) and approximate (for the anharmonic oscillator) calculations. We discuss how this approach could be relevant for the question of interpolation between renormalization group fixed points.

\end{abstract}

\maketitle
\section{Introduction}

The existence of different
behaviors at different scales is a   
central difficulty in quantum field theory.
This is clearly the case for QCD \cite{tomboulis03} 
where the short distance behavior can be described using perturbation 
theory, while confinement and other non-perturbative phenomena
appear at large distance. A 
similar situation is encountered in scalar field theory and spin 
models where, for instance, descriptions of
the renormalization group (RG) flows near the high-temperature fixed point in terms of 
perturbative expansions about the Gaussian fixed point are in general problematic.

In general, the fact that aspects of the RG flows cannot be 
described using perturbation theory is not surprising given that 
perturbative series usually have a zero 
radius of convergence\cite{leguillou90}.
The convergence of perturbative series change drastically \cite{pernice98,convpert} if the large field configurations are removed from the path integral.
This can be implemented, for instance, by removing configurations
such that the norm of local fields exceeds some cutoff value.
The field cutoff modifies the original 
problem, however, it has been found that for nontrivial  
$\phi^4$ problems \cite{convpert} the
modified series apparently converge 
toward values which are exponentially close to the exact ones.
In addition, at fixed order, it is possible in simple examples to choose
the field cutoff in order to optimize the accuracy\cite{optim03,plaquette}.

The calculation of the perturbative series for the modified theories where a large field 
cutoff is introduced is non-trivial. In the case of the anharmonic oscillator, 
approximate analytical methods \cite{tractable} have been developed 
in the limits of large and small field cutoffs. Except at very low order, these methods are not 
accurate in the crossover region between these two regimes and it is imperative to develop new methods that are accurate in the crossover region.
Fortunately, remarkable regularities were observed empirically: with appropriate rescalings and translations, the functions expressing the perturbative coefficients in terms of the 
field cutoff approximately collapse \cite{tractable} along a single curve. 
This suggests that despite the complexity of the perturbative coefficients, 
a simple description of the crossover seems possible.
 
In this article, we refine the description of the approximate collapse of the crossover functions discussed above. An important feature is that the collapse improves as the order increases and a universal (order independent) function is reached asymptotically.  In order to model properly this approach, we start 
with a simpler example, namely the perturbative series associated with a one variable 
integral. This is done in Sec. \ref{sec:int}, where a very good agreement between 
an empirical polynomial parametrization and a saddle point approximation is found.
Generalizations of these results are discussed in Sec. \ref{sec:remarks}.
Our main focus in this article is to discuss the case of the anharmonic oscillator 
``with a field cutoff''. This model is introduced in Sec. \ref{sec:model} and 
recent results regarding the behavior of the perturbative coefficients \cite{tractable} are 
briefly reviewed . Approximate Gaussian parametrizations of the crossover 
are introduced in Sec. \ref{sec:gaussian}. Higher orders corrections are discussed in Sec.
\ref{sec:higher} where a simple description of the crossover for large order emerges.
The possible relevance of this approach for the description of RG flows is discussed in the conclusions. 
\section{The simple integral}
\label{sec:int}

\begin{figure}
\begin{center}
\includegraphics[width=0.6\textwidth]{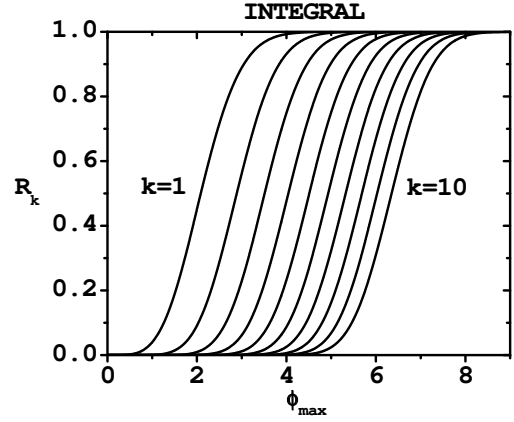}
\caption{The ratio $R_k(x_{max})$
as defined in Eq. (\ref{eq:ratio}). As $k$ goes from 1 to 10, the curves move to the right. } \label{fig:1}
\end{center}
\end{figure}

In this section, we discuss the calculation of the simple integral
\begin{equation}
Z(\lambda )=\int_{-\infty }^{\infty }dx e^{-\frac{1}{2}x
^{2}-\lambda x ^{4}}\ .
\end{equation}
In this example, introducing a field cutoff amounts to remove the 
integration tails 
where $|x |>x _{max}$. Their contributions to the integral is less than $\sqrt{2\pi}{\rm e}^{-\lambda x_{max}^4}$ and if $x _{max}$ is large enough, 
we can approximate $Z(\lambda)$ with 
\begin{equation}
Z(\lambda ,x _{\max })\equiv \int_{-x _{\max }}^{x _{\max
}}dx e^{-\frac{1}{2}x ^{2}-\lambda x ^{4}}\ .
\end{equation}%
If $x _{max} $ is kept at a fixed finite value, the Taylor expansion of the 
exponential converges absolutely and uniformly over the domain of integration and it 
is legitimate to interchange the sum and the integral. The resulting series converges over the entire $\lambda$ complex plane and reads:
\begin{equation}
Z(\lambda ,x _{\max })= \sum_{k=0}^{\infty }a_{k}(x
_{\max })\lambda ^{k}\ ,
\end{equation}%
with 
\begin{equation}
\label{eq:ak}
a_{k}(x _{\max })=\frac{(-1)^{k}}{k!}\int_{-x _{\max
}}^{x _{\max }}dx e^{-\frac{1}{2}x ^{2}}x ^{4k}\ .
\end{equation}
On the other hand, in the limit where $x_{max}\rightarrow\infty$, we have
\begin{equation}
a_{k}(\infty ) 
=\frac{(-1)^{k}}{k!}\Gamma (2k+\frac{1}{2})(2)^{2k+\frac{1}{2}}\ ,
\end{equation}%
and the coefficients grow factorially with the order which implies that the series diverges for any non-zero value of $\lambda$.
In Fig. \ref{fig:1}, we display the ratios
\begin{equation}
\label{eq:ratio}
R_k(x_{max})\equiv \frac{a_{k}(x _{\max
})}{a_{k}(\infty )}
\end{equation}
for $k=1$ to $k=10$. As we can see, the curves corresponding to the different orders have a similar shape and move to 
the right as the order increases. Each curve can be characterized by three regimes: the low field cutoff regime, where the coefficient becomes very small, 
the large field cutoff regime where the coefficient reaches its asymptotic 
value and the crossover region. We now discuss each case separately.
\subsection{The $x _{\max }\rightarrow \infty $ limit}
When $x _{\max }\rightarrow \infty $,
we can use integration by part recursively for the missing tails of 
integration in 
Eq. (\ref{eq:ak}):
\begin{equation}
\label{eq:top}
1-R_k(x_{max})
\simeq e^{-\frac{1}{2}x_{max} ^{2}}\bigg[\frac{(\frac{1}{2}
x _{\max }^{2})^{2k-\frac{1}{2}}}{\Gamma (2k+\frac{1}{2})}+ 
\frac{(\frac{1}{2}x _{\max }^{2})^{2k-\frac{3}{2}}}{\Gamma (2k-\frac{1}{2})}+\cdots
\bigg] \ .
\end{equation}
At each integration by part, the argument of the $\Gamma$ function at the 
denominator decreases by one. Since the arguments are half integers, the 
process does not terminate. Using $1/\Gamma(x-1)=(x-1)/\Gamma(x)$, we end up with a factorial growth for the coefficients of the power series.

\subsection{The $x _{\max }\rightarrow 0 $ limit}
In the opposite limit where $x _{\max }\rightarrow 0 $, 
we expand the exponential in Eq. (\ref{eq:ak}) and we obtain the 
converging series in $x^2_{max}$:
\begin{equation}
R_k(x_{max})
= \frac{2(\frac{1}{2} x _{\max}^{2})^{2k+\frac{1}{2}}}{\Gamma(2k+\frac{1}{2})}\sum_{l=0}^{\infty }\frac{(-\frac{1}{2}x _{\max }^{2})^l}{l!(4k+2l+1)} \ .
\label{eq:bottom}
\end{equation}
It is clear that the factorial growth of the denominator implies that this series converges over the entire $x_{max}$ complex plane.

\begin{figure}
\begin{center}
\includegraphics[width=0.6\textwidth]{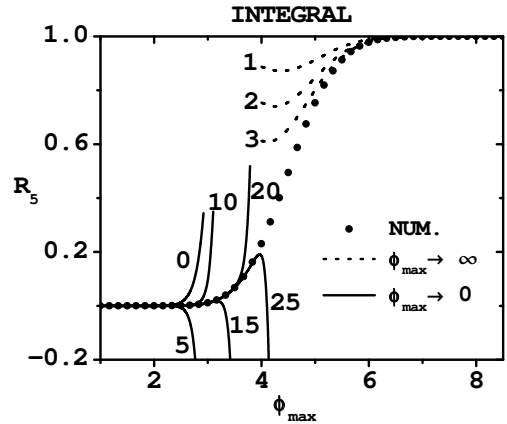}
\caption{The ratio $R_k(x_{max})$
in the large $x _{\max }$ and small $x _{\max
}$ approximations. The dot stands for the accurate numerical
result. The dash lines correspond to the large $x _{\max }$
approximation. The solid lines correspond to small $x
_{\max }$ approximation. The labels $1$, $2$, $3$ correspond to the number of terms in 
Eq. (\ref{eq:top}). The six lines on the 
bottom left correspond to polynomial expansions when $x _{\max
}\rightarrow 0 $ from Eq. (\ref{eq:bottom}) with truncations at order  $0$,
$5$, $10$, $15$, $20$, $25$.
} \label{fig:2}
\end{center}
\end{figure}

\begin{figure}
\begin{center}
\includegraphics[width=0.6\textwidth]{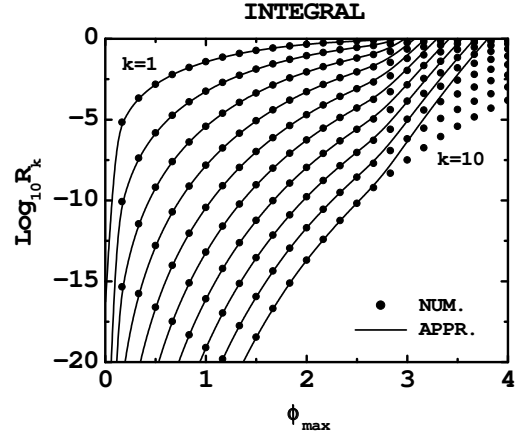}
\caption{The numerical values (dots) of $\log_{10}R_{k}$ and the approximate 
values (lines) obtained with the small $x_{max}$ expansion Eq. (\ref{eq:bottom}) with  10 terms for
$k=1$ to $k=10$} \label{fig:turn-bad}
\end{center}
\end{figure}

In Fig. \ref{fig:2}, we show the curves corresponding to different 
orders in the two expansions discussed above,  
for $k=5$.
The small $x_{max}$ region needs to be resolved logarithmically.
If we restrict the expansion in Eq. (\ref{eq:bottom}) to $l\leq L$, for some given $L$, 
we notice that 
the truncated series becomes a poor approximation when $x_{max}$ reaches some 
critical value (almost independent of the order $k$ in $\lambda$). 
This is illustrated in Fig. \ref{fig:turn-bad} for $L=10$, where the 
critical value is near $x _{\max } \simeq 2.7$.
One can roughly estimate this 
critical value by requiring that the $L$-th term is the same order as the $L-1$-th term.
This yield the critical value $x_{max}\simeq \sqrt{2L}$.  
A more accurate way to proceed is to determine the value of $x_{max}$  for which the $L$-th order approximation becomes poor.
The critical values of $x_{max}$ can be fitted reasonably well with  $0.63+0.64\sqrt{L}$. We can thus conclude that despite the fact that the series is convergent, many terms 
need to be calculated when the order becomes large.
The number of terms that we need to calculate 
in order to get an accurate answer for a given $x_{max}$ grows like 
$x_{max}^2$.

\subsection{Approximate universality}
\label{subsec:appun}

\begin{figure}
\begin{center}
\includegraphics[width=0.7
\textwidth]{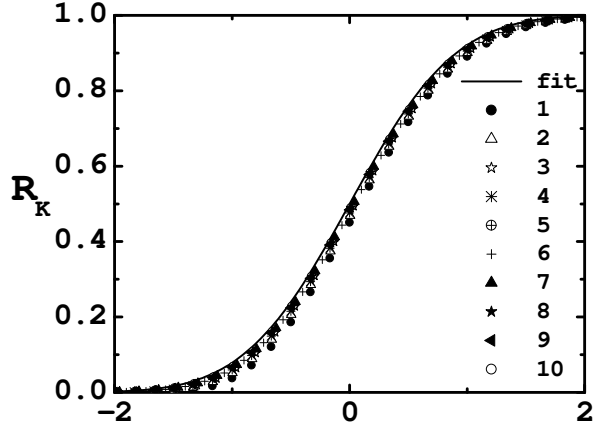}
\caption{Same data as in Fig. \ref{fig:1} but with the argument 
of the $k$-th order curve shifted by $\sqrt{4k}$.
The solid line is $U_{int}$ of Eq. (\ref{eq:gaussapp}).
} \label{fig:intcoll}
\end{center}
\end{figure}

Fig. \ref{fig:2} and the above discussion show that it is not easy to 
use the two expansions in the crossover region, which corresponds to the 
region where the integrand of Eq. (\ref{eq:ak}) peaks.
Using a saddle point approximation about the maximum of the integrand 
$x
=\sqrt{4k}$, we obtain
\begin{equation}
R_k(x_{max}) \simeq
\frac{\int_{0}^{x
_{\max }}dx e^{-2k+4k\ln (\sqrt{4k})}e^{-(x -
\sqrt{4k})^{2}+\cdots}}{\int_{0}^{\infty }dx e^{-2k+4k\ln (
\sqrt{4k})}
e^{-(x -\sqrt{4k})^{2}+\cdots}} \ .
\end{equation}
If we only retain the quadratic term in the above expansion of the phase, we obtain
what we call hereafter the Gaussian approximation. 
This approximation fails to reproduce the small $x_{max}$ behavior
of Eq. (\ref{eq:bottom}). 
It is clear that far from the peak of the integrand, the Gaussian approximation is not good but it is in regions where the original integrand is very small. 
One can simplify this expression by adding the
left tails of integration. This makes sense for $k$ not too small, where it generates 
errors of the order $(4\sqrt{k})^{-1}{\rm e}^{-4k}$.
For sufficiently large $k$, the following approximation holds for $x_{max}$ not to far from $\sqrt{4k}$
\begin{equation}
R_k(x_{max}) \simeq
\pi^{-1/2}\int_{-\infty }^{x _{\max }}dx e^{-(x -\sqrt{4k}%
)^{2}} \ .
\label{eq:apint}
\end{equation}
In Eq. (\ref{eq:apint}), the only dependence on $k$ is in the 
argument of the exponential. Its net effect is a translation in $x_{max}$ and the
shape of the function is universal ($k$-independent). 
In other words, 
\begin{equation}
\label{eq:univ}
R_k(x_{max})\simeq U_{int}(x_{max}-\sqrt{4k}) 
\end{equation}
with 
\begin{equation}
\label{eq:gaussapp}
U_{int}(x)\equiv \frac{\int_{-\infty }^{x }dy e^{-(y^2) 
}}{\sqrt{\pi}} \ .
\end{equation}
If the assumption of universality expressed in Eq. (\ref{eq:univ}) 
is approximately correct, the data should collapse into the universal 
curve when the argument of the $k$-th order curve is shifted by $\sqrt{4k}$. Fig. \ref{fig:intcoll} confirms that this is approximately the case. 
A more detailed look at Fig. \ref{fig:intcoll} indicates that as the order increases, the 
shifted curve gets closer and closer to $U_{int}$. This limit is studied in the next section.

\subsection{Order-dependent corrections}

The basic approximation used to obtain $U_{int}$ in the previous section is the 
saddle point approximation of the integral. We expanded the phase of the integrand 
\begin{equation}
\Phi_k(x_{max})\equiv {\rm ln}R_k'(x_{max})
\end{equation}
about its maximum up to second order. It is possible to improve this approximation by expanding to higher order
\begin{equation}
\hskip-70pt
-\frac{1}{2}x^2+4k {\rm ln}x=2k(2{\rm ln}(2\sqrt{k})-1)-(x-\sqrt{4k})^2-4k\sum_{n=3}^\infty (1/n)\bigg[(\sqrt{4k}-x)/\sqrt{4k}\bigg]^n \ . 
\label{eq:phase}
\end{equation}
The series converges for $0<x<2\sqrt{4k}$. The even terms are negative and if we truncate at even orders, the integral of the exponential of the truncated sum converges when the bounds of integration are sent to infinity. As the order increases, the contribution of the two regions outside the 
radius of convergence are effectively cutoff, because the phase $\Phi_k$ becomes large and negative in these regions. For $x<0$, this is desirable because there is no contribution from this region in the original integral. On the other hand, contributions from $x>2\sqrt{4k}$ exist in the original integral. However, as we move from the maximum of the phase at $\sqrt{4k}$ to the edge of the radius of convergence at $2\sqrt{4k}$, the phase of the integrand drops by $(-6+4{\rm ln}2)k\simeq-3.2k $ and the contribution of the right tail in the original integral is 
small, even for relatively small values of $k$.

\begin{figure}
\begin{center}
\includegraphics[width=0.5\textwidth]{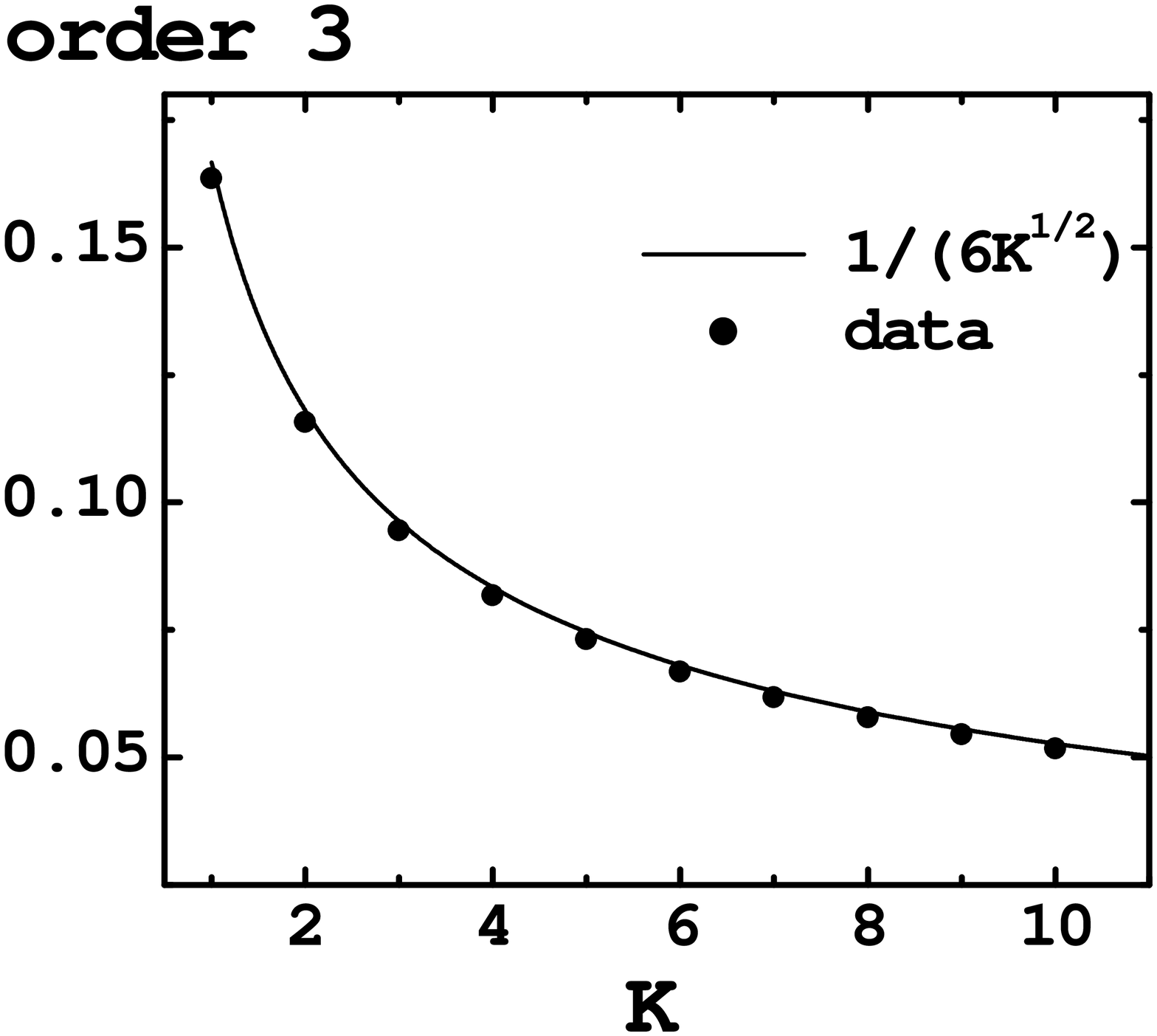}
\includegraphics[width=0.5\textwidth]{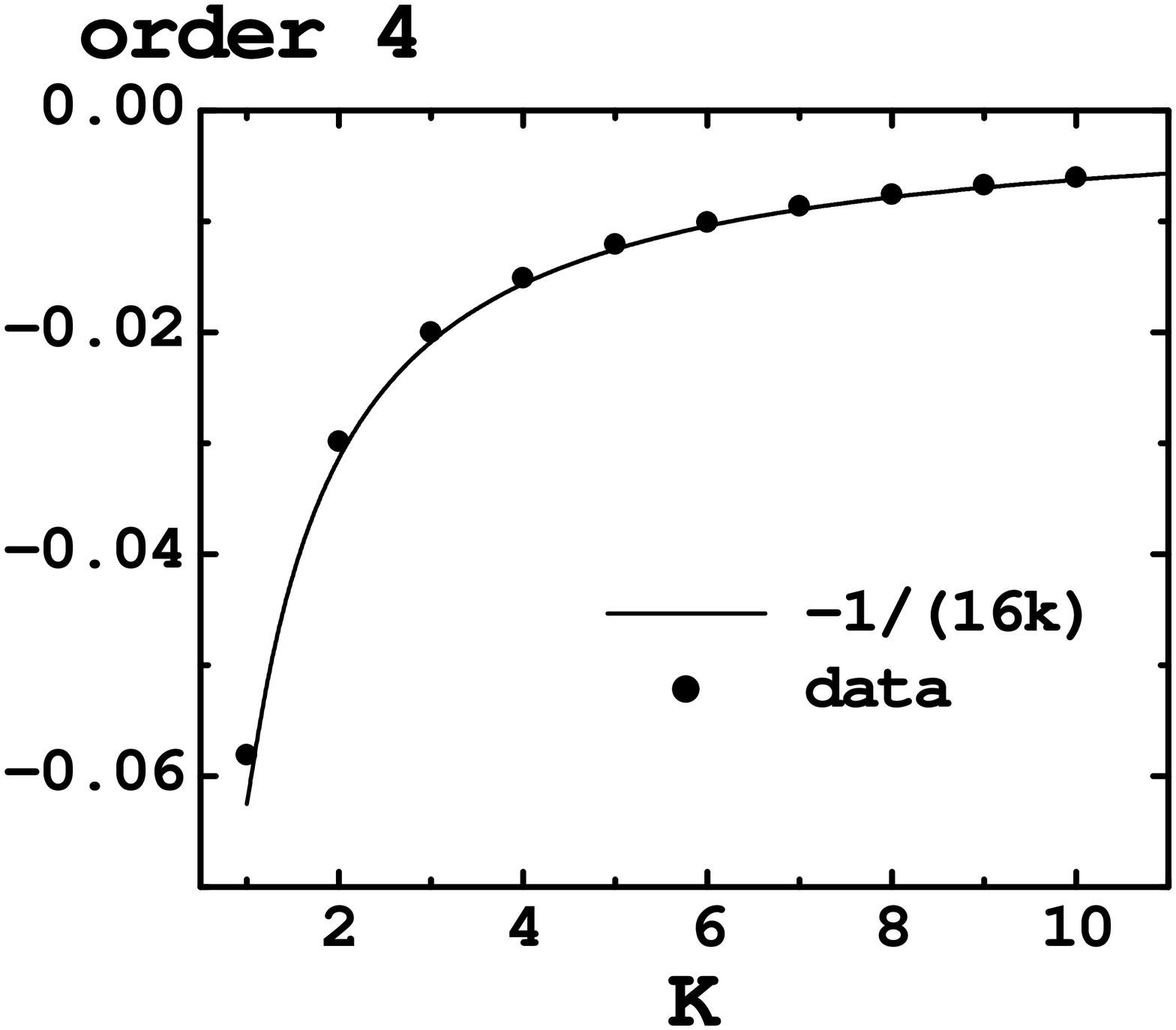}
\caption{Values of the third and fourth coefficients of the Taylor expansion of the phase about its maximum obtained from polynomial fits compared with the predictions of Eq. (\ref{eq:phase}) } \label{fig:int4}
\end{center}
\end{figure}

In order to check the validity of this approach, we have calculated the phase $\Phi_k$ 
by using a discretized approximation of the derivative of $R_k$. We then fitted a set of points near the maximum with polynomials. The results were quite robust under 
changes in the number of points used, the degree of the polynomial and the spacing 
used to calculate the derivative. 
The second coefficient stabilized rapidly between -0.99 and -1 in agreement with 
Eq. (\ref{eq:phase}). The coefficients of order 3 and 4 are shown in Fig. \ref{fig:int4} 
where the good agreement with  Eq. (\ref{eq:phase}) is quite obvious. 

The inclusion of the corrections of order 3 and 4 in the phase 
significantly reduces the small differences between $U_{int}$ and the data in Fig. \ref{fig:intcoll} and, for instance  
for $k=5$, it is impossible to distinguish between the corrected integral and the data with the 
naked eye. 

It is now clear how and why the agreement between the data and the simple guess $U_{int}$ 
of Eq. (\ref{eq:gaussapp}) gets better as the order increases. The size of the crossover region is of order 1 and controlled by the quadratic term in the phase. In this region, 
for $|x-\sqrt{4k}|\sim 1$, $\Phi _k \simeq {\rm const.}- (x-\sqrt{4k})^2$ because corrections of order $n>2$ in Eq. (\ref{eq:phase}) are suppressed by a factor $k^{1-n/2}$. On the other hand the 
argument does not hold far away from the crossover, for $|x-\sqrt{4k}|\sim \sqrt{k}$, where a large number of terms is necessary to approximate logarithms in the phase that 
reproduce the correct power behavior of the integral.

\section{Remarks about the saddle point approximation}
\label{sec:remarks}

In this section, we generalize some aspects of the saddle point approximation used in 
the previous section. 
One important feature of the expansion of the phase given in Eq. (\ref{eq:phase}) is that 
the coefficient of the quadratic term is $k$-independent and twice the value of the coefficient of the quadratic term in the original integral. This has a very simple explanation.
In general, ${\rm e}^{-Ax^2}x^B$ is maximal at $x^\star=\sqrt{B/2A}$ and we have the expansion 
\begin{equation}
	{\rm e}^{-Ax^2}x^B\propto {\rm e}^{-2A(x-x^\star)^2+\dots}\ .
\end{equation}
In doing this calculation, one realizes that in the second derivative of the log estimated at $x^\star$, the factors $B$ cancel exactly.
We want to emphasize that this ``non-renormalization'' is due to the fact that the 
exponential is multiplied by a single power. 
If  we consider instead ${\rm e}^{-Ax^2}(x^B+\epsilon x^{B+ \delta })$, 
then the exact cancellation does not hold.
This can be checked at first order in $\epsilon$, where $x^\star$ is changed according to 
\begin{equation}
\sqrt{B/(2A)}\rightarrow	\sqrt{B/(2A)}\bigg[ 1+(\epsilon\delta/(2B))(B/(2A))^{\delta/2}\bigg]\ ,
\end{equation}
and the coefficient of the quadratic term in the expansion of the phase is changed according to 
\begin{equation}
	-2A\rightarrow-2A\bigg[ 1-(\epsilon\delta^2/(2B))(B/(2A))^{\delta/2}\bigg]\ , 
\end{equation}
and becomes $B$-dependent (except if $\delta =2$ or 0). 
\section{The anharmonic oscillator with a field cutoff}
\label{sec:model}
Our main objective in this article is to understand the crossover for the anharmonic oscillator. 
The model has been studied in detail in Ref. \cite{tractable}. We 
introduce some notations and briefly review some 
basic results.
For convenience, we use quantum mechanical notations 
instead of field theoretical ones. The ``field'' will be denoted $x$ instead of $\phi$ and the field cutoff will be denoted $x_{max}$. In quantum mechanics language, it means that the potential becomes infinite at $\pm x_{max}$. 
We use units such that $\hbar$, $\omega$ and the ``mechanical mass'' $m$ are 1. 
If the dependence on these 3 quantities were restored, 
dimensionless quantities would be expressed in terms of $\sqrt{\omega m/\hbar} x_{max}$ and $\hbar \lambda/m^2\omega^3$. 
The Hamiltonian reads
\begin{equation}
H=\frac{p^{2}}{2}+V(x)
\end{equation}
with
\begin{equation}
V(x)=\left\{
\begin{array}{ccc}
\frac{1}{2}x^{2}+\lambda x^{4}\quad & {\rm if}& |x| < x_{\max}  \\
\infty \quad & {\rm if} & |x| \geq x_{\max} \ .
\end{array}
\right.
\end{equation}
We will consider the perturbative expansion of the ground state:
\begin{equation}
E_0(x_{max}\lambda)=\sum_{k=0}^{\infty }E_0^{(k)}(x_{max})\lambda ^{k}\ .
\label{eq:eexp}
\end{equation}
As in section \ref{sec:int}, we use the notation $R_k$ for the $k$-th coefficient in units of its usual value:
\begin{equation}
R_k(x_{max})\equiv E_0^{(k)}(x_{max})/E_0^{(k)}(\infty) \ .
\end{equation}
We have shown \cite{tractable} that in the limit of small $x_{max}$, 
\begin{equation}
R_k(x_{max})\propto x_{max}^{6k-2}\ ,
\end{equation}
while in the limit of large $x_{max}$, 
\begin{equation}
\label{eq:conj}
1-R_k(x_{\max})\propto
x_{\max }^{4k+1}e^{-x_{\max }^{2}}\ .
\end{equation}

\section{Gaussian approximations of the crossover}
\label{sec:gaussian}

In this section, we discuss Gaussian approximations of the crossover in a way similar to 
what was done in Sec. \ref{subsec:appun} for the integral.
From Eq. (\ref{eq:conj}), we expect that in the large $x_{max}$ limit, 
we have at leading order
\begin{equation}
\label{eq:rp}
R'_k(x_{max})\propto x_{\max }^{4k+2}e^{-x_{\max }^{2}} \ .
\end{equation}
If we now extrapolate backward this behavior at intermediate values of $x_{max}$ 
(this is of course an approximation), $R'_k(x_{max})$  has a maximum at 
\begin{equation}
 x_*(k)\equiv \sqrt{2k+1} \ . 
\end{equation}
Expanding the logarithm up to order 2 about this point, we obtain the Gaussian approximation:
\begin{equation}
\label{eq:g1}
R'_k(x_{max})\propto e^{-2 (x_{\max }- x_*(k))^{2}} \ .
\end{equation}

\begin{figure}
\begin{center}
\includegraphics[width=0.6\textwidth]{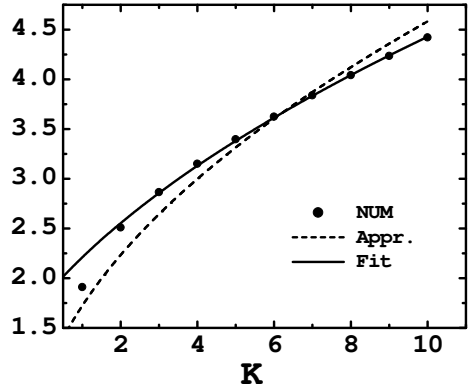}
\caption{The numerical values of $x_0(k)$ (circles) compared to $\sqrt{2k+1}$ (dash line) and $\sqrt{1.63k +3.26}$ (solid line). }
\label{fig:shift-anh}
\end{center}
\end{figure}

With this approximation, we have $R''_k(x_*(k))=0$. This approximate 
prediction for the value of $x$ for which the second derivative vanishes can be compared with the 
accurate numerical value, denoted $x_0(k)$,  for which it actually 
vanishes. The numerical values can be found in Table 2 of Ref. \cite{tractable} 
and are shown in Fig. \ref{fig:shift-anh}, together with approximate analytical formulas. Typically, $x_0(k)\simeq x_*(k)$ with a 10 percent accuracy for $k\leq 10$. A more accurate formula can be found by fitting $x_0(k)^2$ with a linear function. 
Using the data for $k=2$ to 10, we obtain $x_0(k)\simeq \sqrt{1.63k +3.26}$.
Using the data for $k=10$ to 20, we obtain a slightly different fit that we expect to 
be closer to the asymptotic behavior:
\begin{equation}
	x_0(k)\simeq \sqrt{1.61k +3.48}\ .
	\label{eq:appx0}
\end{equation}

\begin{figure}
\begin{center}
\includegraphics[width=0.8\textwidth]{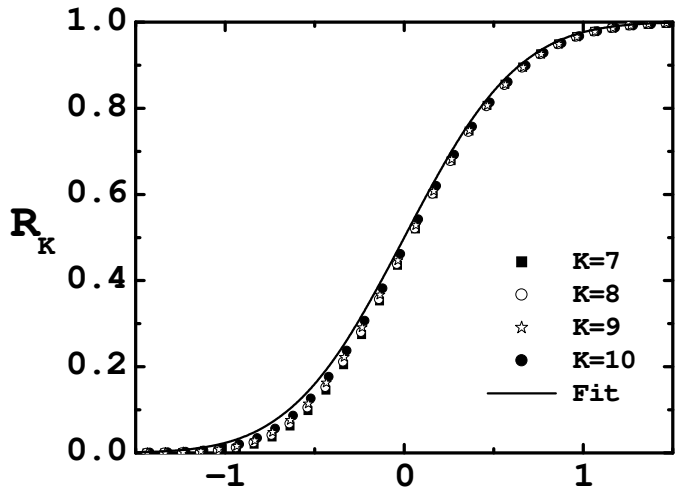}
\caption{$R_k(x_+x_0(k))$ for $k=7, \dots 10$ and the function 
$U_{anh,1}(x)$ of Eq. (\ref{eq:propu}). }
\label{fig:fit-an}
\end{center}
\end{figure}

In Ref. \cite{tractable}, we found that if $R_k(x_{max})$ is translated by 
$x_0(k)$, the data for $k=2,\dots 10$ approximately collapses.
In other words,
\begin{equation}
	R_k(x+x_0(k))
	\simeq U_{anh}(x)\ .
	\end{equation}
An approximate form for $U_{anh}(x_{max})$ is suggested by Eq. (\ref{eq:g1}). 
Integrating and normalizing by requiring $lim_{x\rightarrow +\infty}\ U_{anh}(x)=1$, we obtain	
\begin{equation}
U_{anh, 1}(x)
\label{eq:propu}
\simeq \sqrt{2/\pi}\int_{-\infty }^{x}dy {\rm e}^{-2y^2} \ .
\end{equation}
We have used the subscript $anh,1$ to indicate that it is a first approximation and that 
a better form will be found later. 
This proposal is compared with the empirical data 
for $k=7,\ \dots ,10$ in Fig. \ref{fig:fit-an}. 
We only displayed a few values of $k$ so that it is possible to distinguish the various orders on the graph.
One can see that the collapse of the various curves is reasonably good. The function 
$U_{anh,1}$ fits the collapsed data quite well on the upper right, but appears to be slightly above the data on the lower left. However, as the order increases (in the Figure from 7 to 10), the difference diminishes.

The reason why $\sqrt{2k+1}$ is not a very good approximation of $x_0(k)$ is that the large $x_{max}$ behavior given by Eq. (\ref{eq:rp}) is not accurate for intermediate values of $x_{max}$. Studying numerically the behavior of 
\begin{equation}
R_k(x_{max}){\rm e}^{x_{max}^2} \ ,
\label{eq:rkm}
\end{equation}
near $x_0(k)$, ones obtains a very good linear behavior on a log-log plot.
If Eq. (\ref{eq:rp}) were correct, the slope would be $4k+2$. However, 
collecting the slopes for various values of $k$, one concludes that $4k+2$  should be replaced by $3.2k+7.8$ in Eq. (\ref{eq:rp}). This is consistent with the empirical form of $x_0(k)$ 
of Eq. (\ref{eq:appx0}) which is approximately the square root of one half of this fit
(see Sec. \ref{sec:remarks} with $A$=1.)
When studying the difference between Eq. (\ref{eq:rkm}) and its approximation 
by a single power, one notices small quadratic corrections. 
We will not report all the details of this investigation, but only mention that 
these quadratic corrections 
can be approximately removed by replacing Eq. (\ref{eq:rkm}) by $R'_k(x_{max}){\rm e}^{\beta _k x_{max}^2}$ and adjusting the value of $\beta_k$ to the value for which the quadratic corrections change sign. 

We conclude that a more realistic version of Eq. (\ref{eq:rp}) is 
\begin{equation}
\label{eq:rpimpr}
R'_k(x_{max})\propto x_{\max }^{3.2k+7.8}e^{-\beta_k x_{\max }^{2}} \ ,
\end{equation}
The results of Sec. \ref{sec:remarks} suggest that ${\rm e}^{-2 y^2}$ should be replaced by ${\rm e}^{-2\beta_k y^2}$ in Eq. (\ref{eq:propu}). This brings an explicit dependence on $k$, and universality can only be reached asymptotically if $\beta_k$ tends to a limit 
when $k$ becomes large.

Empirically, the values of $\beta_k$ are slightly larger than 1 for $k<10$ and slightly smaller than 1 for $k>10$. This modification allows to reduce 
the small discrepancies of the Gaussian approximation used before. Note that in Fig. 
\ref{fig:fit-an}, the closeness of $k=9$ and 10 to $U_{anh,1}$ reflects the fact that 
$\beta_9$ and $\beta_{10}$ are very close to 1. $\beta_k$ is closely related to 
$A_{k}^{(2)}/2$, a quantity that will be studied in the next section. Anticipating the results presented 
there, we expect that as $k$ increases, $\beta_k$ stabilizes to a value close to 
0.75. 

The limitations of the Gaussian approximation, or of any other expression of 
a universal function $U$ as the integral of a even function, can be seen from 
the duality relation which exchanges the small and large field cutoff regions
\begin{equation}
	U(-x)=1-U(x)\ ,
	\label{eq:dual}
\end{equation}
which cannot be exact since the approach of 0 when $x_{max}\rightarrow 0$ is power like 
while in the large $x_{max}$ limit, the approach of 1 is exponentially small.
In at least one of these two limits, $U$ should fail to provide the correct behavior. 

\section{Numerical study of the higher order corrections}
\label{sec:higher}

\begin{figure}
\begin{center}
\includegraphics[width=0.5\textwidth]{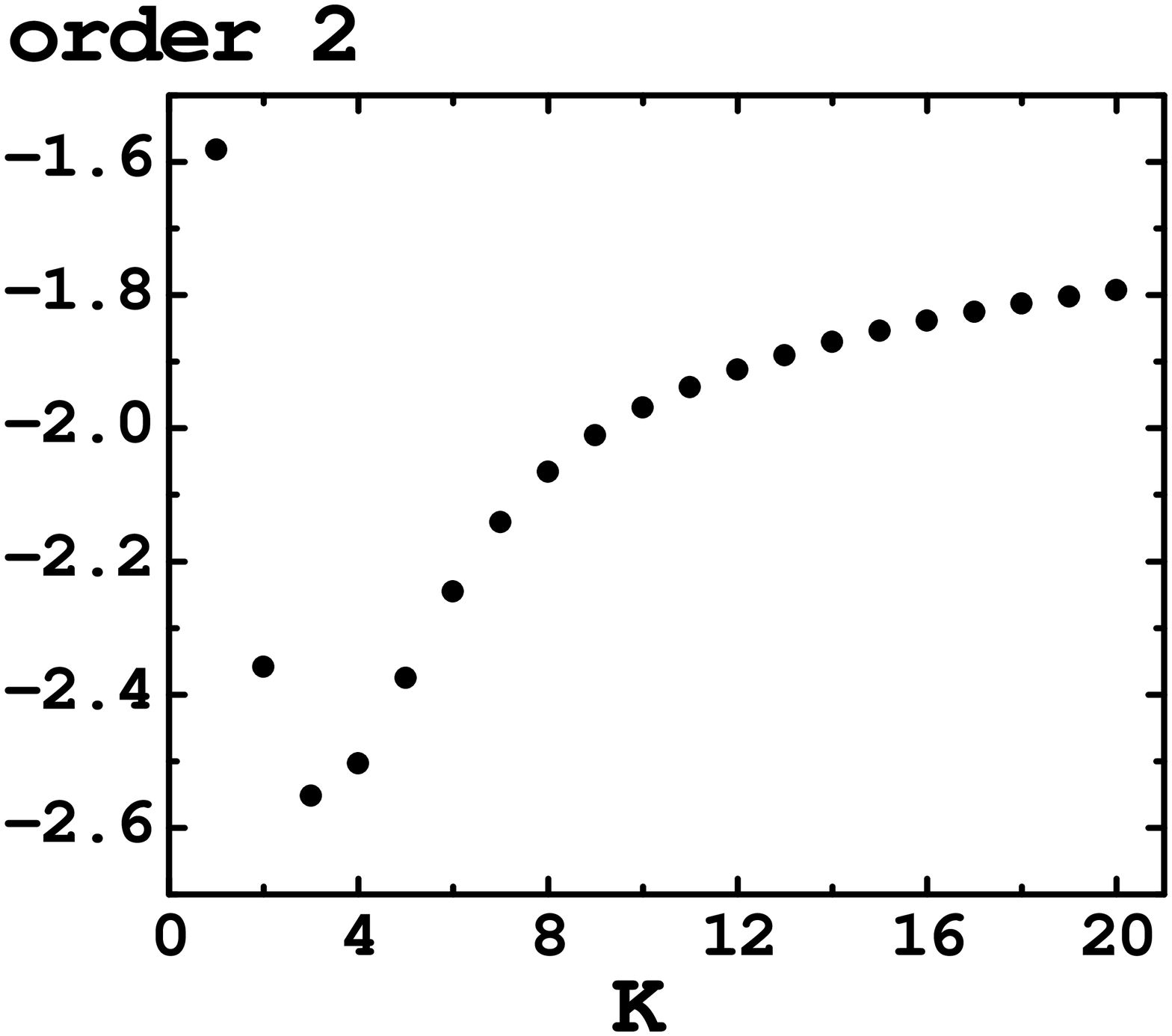}
\includegraphics[width=0.5\textwidth]{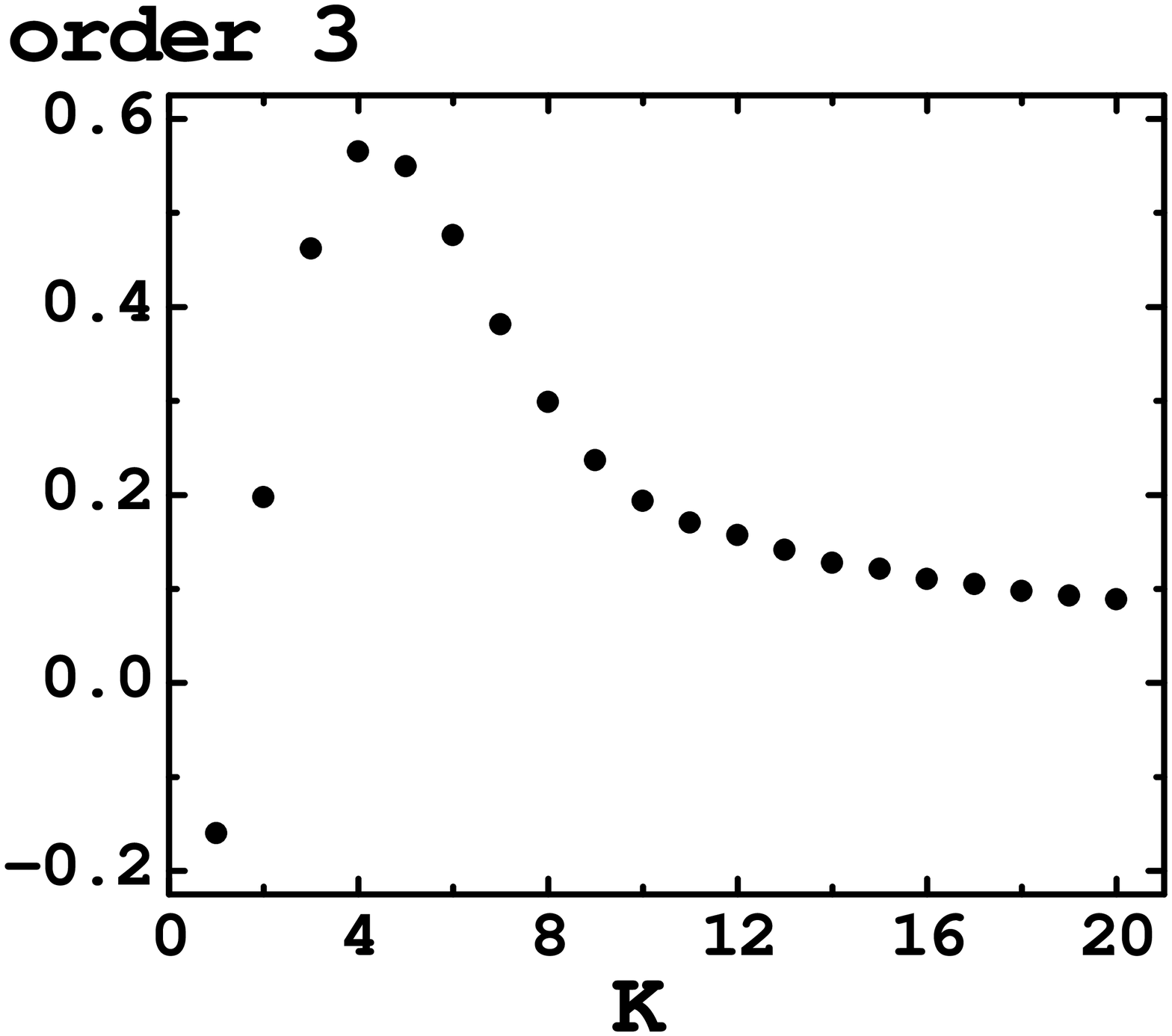}
\includegraphics[width=0.5\textwidth]{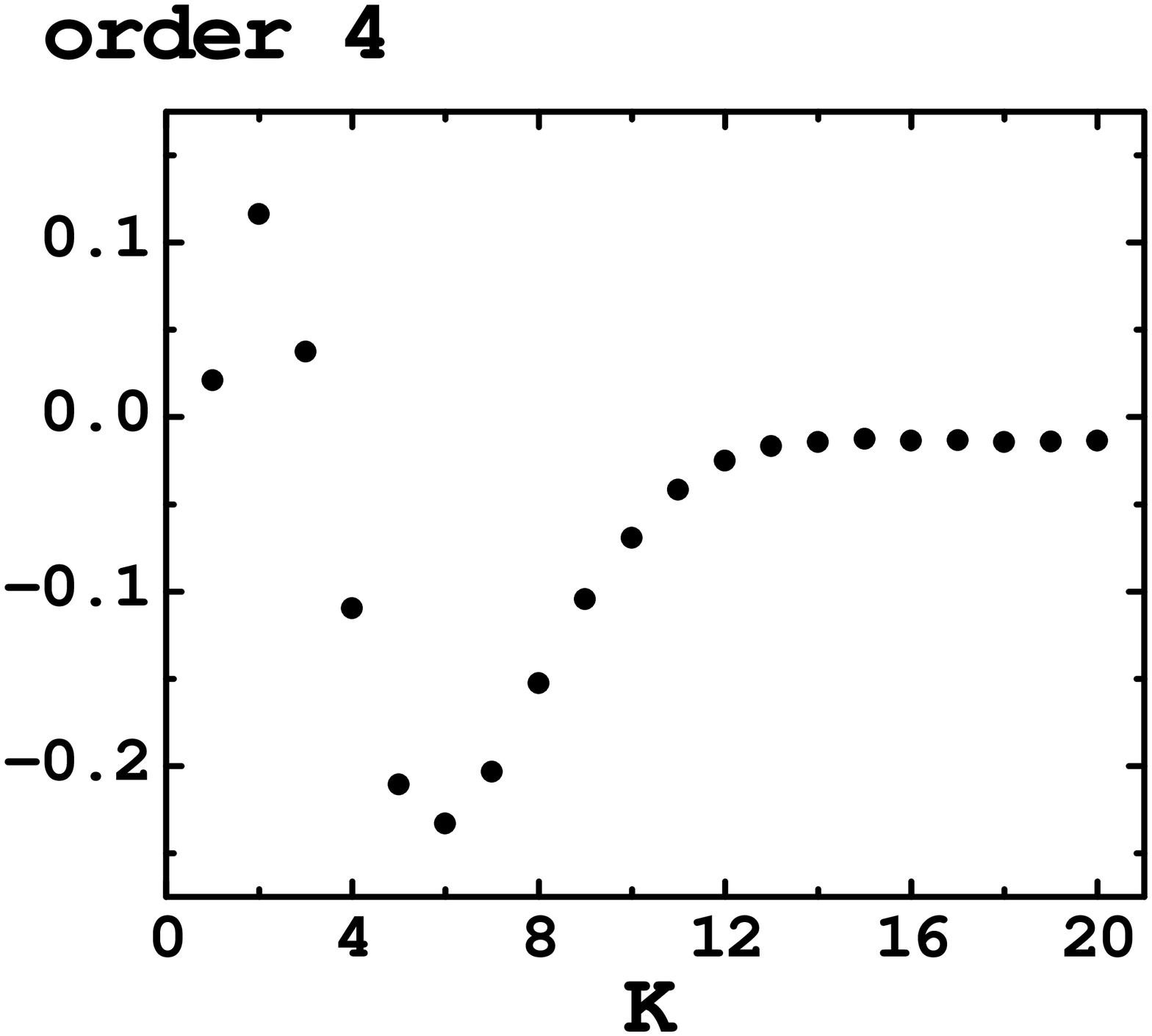}
\caption{$A^{(q)}_k$ for $q=2$, 3 and 4 and $k= 1, \ \dots 20$. } \label{fig:anh2}
\end{center}
\end{figure}

The complexity of the exact form of the terms of the perturbative series increases 
rapidly with the order (see for instance \cite{simonIV} for exact expression up to order 4).
The number of graphs grows factorially \cite{bender76}. It seems unlikely, that 
at this point, 
one could guess approximate equations such as Eq. (\ref{eq:rpimpr}), or an improved 
version of it, from analytical expressions at large order. 
For this reason, we will describe ${\rm ln}(R'_k)$ near $x_0(k)$ using 
polynomial approximations. This method is numerically robust and has been tested 
successfully in the case of the integral where the expansion can be calculated 
analytically. Our main working assumption is that in the vicinity of $x_0(k)$:
\begin{equation}
{\rm ln}(R'_k(x)\simeq  \sum_{q=0}^Q{A^{(q)}_k(x-x_{0}(k))^{q}}
\end{equation}
Note that the presence of odd powers in the expansion breaks the duality expressed in 
Eq. (\ref{eq:dual}). 
The fitted value for the largest order ($q=Q$) coefficient is typically not very stable and as we want to focus on the 
terms with  $q=2$, 3 and 4,
we have used $Q=5$ and 6. We then checked that the terms $A^{(q)}_k$ for $q=2$, 3 and 4 were quite stable under small changes as in Sec. \ref{sec:int}. They are displayed in Fig. \ref{fig:anh2}. 

As the  $A^{(q)}_k$ seem to reach asymptotic values, we have fitted the larger $k$ values with 
the parametric form
\begin{equation}
A_k^{(q)}=a^{(q)}+b^{(q)}k^{-c^{(q)}}
\end{equation}
Other parametric forms such as an exponential decay have been tried too but have been shown 
to have a significantly larger chi-square.
We first estimated $c^{(q)}$ by making a log-log plot of $|A_{k+1}^{(q)}-A_k^{(q)}|$ and 
then obtained $a^{(q)}$ and $b^{(q)}$ with standard linear fit methods. We used different sets of $A^{(q)}_k$ with $k$
running from $k_0$ to 20. The results showed a regular dependence on $k_0$: as $k_0$ 
increased, the three fitted parameters showed small, monotonous changes and we then attempted 
an extrapolation (to large $k_0$) using the same method as above. This 
second extrapolation is of course much more delicate and used only to estimate the errors. 
The stability of the procedure was checked by using five different sets of $A_k^{(q)}$ 
obtained with slightly different procedures: changes in the range of $x_{max}$ 
used in the fits, changes in the $\Delta x_{max}$ used to calculate the derivative of $R_k$, changes in $Q=5$ or 6.

For $q$=2, the fits at fixed $k_0$ were found to be quite consistent among the five sets 
and the second extrapolation was in some cases possible. Our final estimate is 
\begin{equation}
A_k^{(2)}\simeq-1.5(1)+2.0(1)k^{-0.6(1)} \ ,
\end{equation}
with the parenthesis indicating the estimated error on the last written digit.
For sufficiently large $k$, we found that $A_k^{(2)}\simeq -2\beta_k$ as defined in the 
previous section. It is clear form Fig. \ref{fig:anh2} that for $k\simeq 10$, $A_k^{(2)}\simeq -2$ in agreement with the observation $\beta_{10}\simeq 1$. 
To give an idea about the accuracy of the agreement, for $k=15$, $A_{15}^{(2)}\simeq -1.866$ while $\beta_{15}=0.938=1.876/2$. 

For $q=3$, more variations among the five data sets were observed. A tentative value for $a^{(3)}$ is 0.02(1). However this is more than 4 time less than $A_{20}^{(3)}\simeq 0.089(1)$. More stable results with consistent extrapolations were obtained by setting 
$a^{(3)}=0$. Our final estimate is 
\begin{equation}
A_k^{(3)}\simeq 1.3(3)k^{-0.9(1)} \ .
\end{equation}
We repeat that it is difficult to rule out a small positive value asymptotically.
The decay rate seems faster than in the single power case studied in Sec. \ref{sec:remarks} where $A_k^{(3)}\propto k^{-1/2}$. In our fits the average value of the 
power was -0.94 for $k_0=17$ and the extrapolated value the farther away from this average was -0.86.

For $q=4$, stronger variations among the five sets were observed. 
In all cases $a^{(4)}$ was much smaller than $A_{20}^{(4)}\simeq -0.017(1)$ and with no 
consistent sign.
More stable results with reasonably consistent extrapolations were obtained by setting 
$a^{(4)}=0$. Our final estimate is 
\begin{equation}
A_k^{(4)}\simeq -1.5(6)k^{-1.3(3)}
\end{equation}

We have also considered the coefficients for larger values of $q$ and found that they also
appear to reach zero asymptotically. Their values are in general quite small. 
For instance, $A_{20}^{(5)}\simeq 0.004$ and $A_{20}^{(6)}\simeq -0.0007$. 

We conclude that for $k$ sufficiently large and $x_{max}=x_0(k)+x$, with $x$ of order 1,  
we have in good approximation that 
\begin{equation}
	R_k(x_0(k)+x)
	\simeq U_{anh}(x)
\simeq \sqrt{2A/\pi}\int_{-\infty }^{x}dy {\rm e}^{-2Ay^2} \ , 
\end{equation}
with $A\simeq 0.75$. The words of caution regarding this ``limit'' are the same as in the 
case of the integral: if $|x|$ is too large, the non-Gaussian corrections need to be taken 
into account.

\section{Conclusions}
In summary, we have considered the simplest field theoretical models 
with a field cutoff, in zero and one dimension. We found in both cases, that the crossover behavior of the perturbative coefficient can, for orders large enough, be 
described in good approximation by rescaling and translating a universal (order-independent) function.
Furthermore, this function can be expressed as a Gaussian integral. 
This statement is exact for the simple integral (in the sense of Eq. (\ref{eq:phase})). 
For the anharmonic oscillator, small non-Gaussian corrections persisting at large order 
cannot be ruled out.

We expect similar features in higher dimensional scalar field theory with a UV 
regulator. Generically, a modified perturbative calculation involves a few 
coefficients with values close to their usual ones, a few coefficients in the crossover region, and the rest of the coefficients taking values much smaller than 
in regular perturbation theory. These three regimes are reminiscent of the three regimes encountered when calculating renormalization group flows between two fixed points. One should notice the similarities between the graphs presented here and those of Refs. \cite{bagnuls01,pelissetto98} where this type of crossover behavior is studied.
In the same way, the approximate duality of Eq. (\ref{eq:dual}) reminds of the 
duality between fixed points of a simplified renormalization group equation \cite{dual} or the exchange between the perturbative and non-perturbative sectors 
of quasi-exactly solvable periodic potentials
\cite{dunne02}. 

We plan to make these analogies more specific  
(or disprove these ideas), by 
making explicit calculations with Dyson's hierarchical model. 
Namely, we plan to compare the relative weights of the three types of 
perturbative terms discussed above with the relative increases
in the three regions of the renormalization group flows discussed 
above for the zero-momentum $n$-point functions.
Our hope is that we will be able to trade the very complicated problem of the 
interpolation between RG fixed points to a more tractable one.

\ack
This research was supported in part by the Department of Energy
under Contract No. FG02-91ER40664 and also by
the Obermann Center for Advanced Studies at the
University of Iowa. 

\section*{References}


\end{document}